\documentclass[12pt]{iopart}

\usepackage{graphicx}

\newcommand{\cfigl}[3]{\begin{figure}[!hbtp]\centering
 \includegraphics[width=14cm]{#2}\caption{\small{{#3}}}\label{#1}\end{figure}}

\usepackage{iopams}
\newcommand{\field}[1]{\mathbb{#1}}
\newcommand{\RR}{\field{R}}

\begin{document}

\title[Is GR simpler?]
{Is General Relativity a simpler theory?}

\author{Mart\'{\i}n Rivas}
\address{Theoretical Physics Department, University of the Basque Country,\\ 
Apdo.~644, 48080 Bilbao, Spain}
\ead{martin.rivas@ehu.es}

\begin{abstract}
Gravity is understood as a geometrization of spacetime. But spacetime is also the manifold 
of the boundary values of the spinless point particle in a variational approach. Since all known
matter, baryons, leptons and gauge bosons are spinning objects, it means that the manifold, which we call the kinematical
space, where we play the game of the variational formalism of an elementary particle 
is greater than spacetime. 
This manifold for any mechanical system is a Finsler metric space such that 
the variational formalism can always be interpreted
as a geodesic problem on this space. 
This manifold is just the flat Minkowski space for the free spinless particle.
Any interaction modifies its flat Finsler metric as gravitation does. 
The same thing happens for the spinning objects
but now the Finsler metric space has more dimensions and its metric 
is modified by any interaction, so that
to reduce gravity to the modification only of the spacetime metric is to make a simpler theory, 
the gravitational theory of spinless matter. Even the usual assumption that the modification
of the metric only involves dependence of the metric coefficients on the spacetime variables is also
a restriction because in general these coefficients are dependent on the velocities. In the spirit of unification
of all forces, gravity cannot produce, in principle, a different and simpler geometrization than any other interaction.

\end{abstract}

%

\maketitle

\section{Introduction}
\label{sec:intro}
{\sl Things should be made simple, but not simpler}. From this sentence attributed to Albert Einstein
is where we take the title of this work to show that if the spin concept
of elementary particles had been known to physics before General Relativity was born
most probably the geometrization of spacetime proposed by its creator
should be changed by the geometrization of a different manifold, larger than spacetime,
so that today's General Relativity would be considered as a theory of gravitation 
of simpler and spinless matter.

The variational approach of classical mechanics can always be interpreted as a geodesic statement
on the space $X$ of the boundary variables of the variational formalism \cite{Rivasbook}. 
But this metric manifold
$X$, is not a pseudo-Riemannian space but rather a Finsler space \cite{Asanov}, \cite{Rund},
where the symmetric metric tensor $g_{ij}(x,\dot{x})$ is not only a function of the point $x\in X$, 
but also of its velocity $\dot{x}$, where the
overdot means derivative with respect to some arbitrary evolution parameter. For the relativistic 
spinless or point particle
this manifold $X$ is just the spacetime $\RR^4$ and in the free case the metric is Minkowski's 
metric $\eta_{\mu\nu}$. But if the particle has spin it would have more degrees of freedom, 
so that the variational approach will be described as a geodesic problem in a larger manifold than spacetime. 
Interactions and gravitation would modify the metric of this larger
manifold, so that to restrict ourselves to the geometrization of the spacetime submanifold is to simplify
the problem, or in physical terms, to reduce the gravitational behaviour of real spinning matter 
to that of spinless and unexistent matter.

In the next section \ref{sec:geodesic}, I will make a summary of the variational approach 
of classical mechanics, which shows how it can be interpreted as a geodesic statement 
and the way to obtain the metric from the Lagrangian. 
If the Lagrangian of a free elementary particle is modified by any interaction, the metric is modified.
Any interaction modifies the geometry of the $X$ manifold.

In section \ref{sec:metrics} we analyze some examples
to show the Finsler metric structure of the
spacetime of a charged point particle under some external interactions, which include a uniform magnetic field
and a constant gravitational field and the Newtonian gravitational field of a point mass $M$. In all cases
the modification of the metric coefficients involve dependence on the velocities of the point particle.
A Riemannian approximation to the metric can be obtained in the low velocity limit.

In section \ref{sec:spin} after a short introduction to the concept of classical elementary particle
I will describe the most general $X$ manifold of a relativistic spinning particle which satisfies Dirac's
equation when quantized. It is on this manifold that the plausible generalization of 
Einstein's formalism has to be worked out.

As a general conclusion, General Relativity is a restricted theory of gravitation 
and therefore a simpler theory in two aspects.
One, that the manifold whose geometry is changed by any interaction is larger than spacetime because 
elementary particles are spinning particles. The second is that the modification of the metric coefficients
should involve dependence on the velocities.
 
\section{The geodesic interpretation of the variational formalism}
\label{sec:geodesic}

Let us consider any mechanical system of $n$ degrees of freedom $q_i$, $i=1,\ldots,n$, 
described by a Lagrangian, $L(t,q_i,q_i^{(1)})$, where $t$ is the time and $q_i^{(1)}=dq_i/dt$.
The variational approach is stated in such a way that the path followed by the system makes
stationary the action functional 
\[
{\cal A}[q(t)]=\int_{t_1}^{t_2}L(t,q_i,q_i^{(1)})dt,
\]
between the initial state $x_1\equiv(t_1,q_i(t_1))$ and final state $x_2\equiv(t_2,q_i(t_2))$
on the $X$ manifold, which is the $(n+1)$-th dimensional manifold spanned by the time $t$ and the $n$ degrees of freedom
$q_i$.
If instead of describing the evolution in terms of time we express the evolution in parametric
form $t(\tau)$, $q_i(\tau)$ in terms of some arbitrary evolution parameter $\tau$, then
$q_i^{(1)}(\tau)=\dot{q_i}/\dot{t}$, where now the overdot means $\tau$-derivative.
The variational approach will be written as
\[
\int_{\tau_1}^{\tau_2}L(t,q_i,\dot{q_i}/\dot{t})\dot{t}d\tau=\int_{\tau_1}^{\tau_2}\widetilde{L}(x,\dot{x})d\tau,\quad \widetilde{L}=L\dot{t},
\]
with the same boundary values on the $X$ manifold as before $x_1$ and $x_2$. 
But now the Lagrangian $\widetilde{L}$ is independent
of the evolution parameter $\tau$ and it is a homogeneous function of first degree of the derivatives $\dot{x}$ \cite{Rund}.
This means that $\widetilde{L}^2$ is a positive definite homogeneous function of second degree of the derivatives $\dot{x}$, so that
Euler's theorem on homogeneous functions allows us to write
\[
\widetilde{L}^2=g_{ij}(x,\dot{x})\dot{x}^i\dot{x}^j,\quad i,j=0,1,\ldots n
\]
where index 0 corresponds to the time variable and the $g_{ij}$ are computed from $\widetilde{L}^2$ by
\[
g_{ij}(x,\dot{x})=\frac{1}{2}\frac{\partial^2\widetilde{L}^2}{\partial\dot{x}^i\partial\dot{x}^j}=g_{ji}.
\]
Between the allowed boundary states $x_1$ and $x_2$, since $\widetilde{L}^2>0$, 
the metric $g_{ij}(x,\dot{x})$ represents a definite
positive metric which transforms as a second rank covariant tensor under transformations which leave 
$\widetilde{L}$ invariant \cite{Asanov,Rund}. The variational problem can be stated as
\[
\int_{\tau_1}^{\tau_2}\widetilde{L}(x,\dot{x})d\tau=\int_{\tau_1}^{\tau_2}\sqrt{\widetilde{L}^2(x,\dot{x})}d\tau=
\int_{\tau_1}^{\tau_2}\sqrt{g_{ij}(x,\dot{x})\dot{x}^i\dot{x}^j}d\tau=
\]
\[
=\int_{x_1}^{x_2}\sqrt{g_{ij}(x,\dot{x})d{x}^id{x}^j}=\int_{x_1}^{x_2}ds,
\]
where $ds$ is the arc length on the $X$ manifold with respect to the metric $g_{ij}$. The variational statement
has been transformed into a geodesic problem with a Finsler metric. As shown in \cite{Rivasbook} this is valid even
for Lagrangian systems depending on higher order derivatives $L(t,q_i,q_i^{(1)},\ldots,q_i^{(k)})$, 
$q_i^{(k)}=d^kq_i/dt^k$. In this case the manifold of the boundary variables $X$, which will be called the 
kinematical space from now on, is spanned by the time $t$, the $n$ degrees of freedom $q_i$ and their corresponding time derivatives
up to order $k-1$.

Since $\widetilde{L}$ is homogeneous of first degree in terms of the derivatives $\dot{x}$ can also be decomposed
as a sum of terms with dimensions of action if the arbitrary evolution parameter is taken dimensionless,
\[
\widetilde{L}=\frac{\partial\widetilde{L}}{\partial\dot{x}^i}\dot{x}^i=F_i(x,\dot{x})\dot{x}^i,
\]
where the $F_i(x,\dot{x})$ are homogeneous functions of zero-th degree of the $\dot{x}^i$, so that they involve
time derivatives of the different variables. The metric coefficients can be expressed as
 \begin{equation}
g_{ij}=F_iF_j+\widetilde{L}\frac{\partial^2\widetilde{L}}{\partial\dot{x}^i\partial\dot{x}^j}=F_iF_j+\widetilde{L}\frac{\partial F_i}{\partial\dot{x}^j}=g_{ji}
 \label{eq:metrica}
 \end{equation}
and are also homogeneous functions of zero-th degree of the $\dot{x}^i$. 

As an example, the relativistic point particle of mass $m$ and spin 0 has a kinematical space
spanned by time $t$ and the position of the point ${\bi r}$, so that the free Lagrangian
$\widetilde{L}_0=\pm mc\sqrt{c^2\dot{t}^2-\dot{\bi r}^2}$, ($+$ for the antiparticle, $-$ for the particle) 
is clearly a homogeneous function of first
degree of the derivatives $\dot{t}$ and $\dot{\bi r}$. With $x^0\equiv ct$, the Finsler metric becomes
\[
F_\mu=\frac{\partial\widetilde{L}_0}{\partial\dot{x}^\mu}=-p_\mu=\mp\frac{mc\dot{x}_\mu}{\sqrt{\dot{x}_\nu\dot{x}^\nu}},
\qquad g_{\mu\nu}=p_\mu p_\nu-\widetilde{L}_0\frac{\partial p_\mu}{\partial\dot{x}^\nu}=m^2c^2\eta_{\mu\nu},
\] 
where $\eta_{\mu\nu}$ is diag$(1,-1,-1,-1)$ and $p_\mu$ the energy-momentum four-vector. 
The interaction with some
external electromagnetic field is described by the new Lagrangian $\widetilde{L}=\widetilde{L}_0+\widetilde{L}_I$, with
$\widetilde{L}_I=-eA_\mu(x)\dot{x}^\mu$, so that the variational problem is transformed into a geodesic problem with 
a new metric on $X$ space, given by
\[
F_\mu=\frac{\partial\widetilde{L}}{\partial\dot{x}^\mu}=-p_\mu-eA_\mu,\]
 \begin{equation}
g_{\mu\nu}(x,\dot{x})=m^2c^2\eta_{\mu\nu}+e^2A_\mu A_\nu+e(p_\mu A_\nu+p_\nu A_\mu)
+eA_\sigma\dot{x}^\sigma\frac{\partial p_\mu}{\partial\dot{x}^\nu}.
 \label{eq:metricaEM}
 \end{equation} 
The modification of the metric vanishes when $e\to 0$. Because $p_\mu$ is not explicitely dependent
of the variables $x$, the dependence of the metric on the spacetime coordinates is coming only from the 
external fields $A_\mu(x)$. But it depends on the $\dot{x}$ variables through its dependence
on the $p_\mu$ and its derivatives. 

\section{Examples of Finsler spaces}
\label{sec:metrics}

In figure \ref{fig:motions} we show three possible motions of a charged point particle in its kinematical space,
which reduces in this case to the spacetime. 
The three trajectories are geodesics of spacetime but with respect to three
different metrics. In part (a) the motion is free,
the trajectory is a straight line. In (b) the particle is under the action
of an external uniform magnetic field, and the trajectory has curvature and torsion.
In this case the Finsler metric of spacetime is different than the metric in the free case.
The external magnetic field modifies the metric. Finally in (c) it is the same free trajectory but as seen
by an accelerated observer. According to the equivalence principle this is equivalent
to the analysis under a global and constant gravitational field. Also in this case the metric has been modified.

\cfigl{fig:motions}{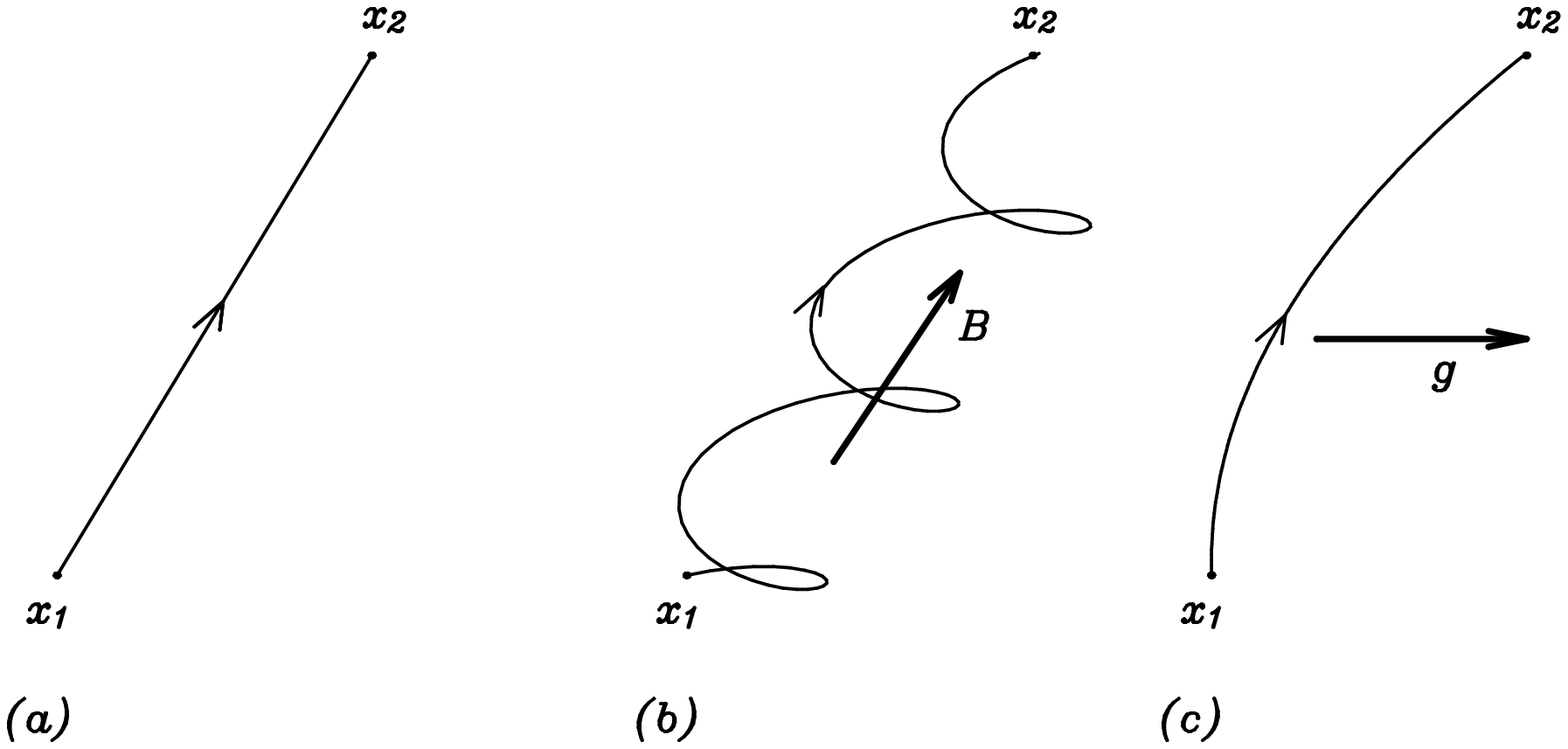}{Three motions of a point particle in its kinematical space
between the boundary points $x_1$ and $x_2$. (a) in the free case, (b) under a uniform magnetic field ${\bi B}$, and 
(c) under a uniform gravitational field ${\bi g}$. In the three cases the kinematical space
is the same, the spacetime, the trajectories are geodesics but with respect to three different Finslerian metrics. The spatial part of the trajectories is
in the case (a) a straight line with no curvature and no torsion, in
(b) with curvature and torsion and in (c) a flat trajectory with curvature.}

In case (a) the metric is $g_{\mu\nu}=m^2c^2\eta_{\mu\nu}$, where $\eta_{\mu\nu}$ is diag$(1,-1,-1,-1)$. 
It is the constant Minkowski metric.

In the case (b), let us assume a uniform magnetic field along $OZ$ axis of intensity $B$. 
We can take as the potential vector ${\bi A}=(0,Bx,0)$ and scalar potential $A_0=0$. 
The Lagrangian of the point particle under this field is
 \begin{equation}
\widetilde{L}_B=-mc\sqrt{\dot{x}_0^2-\dot{\bi r}^2}+eBx\dot{y}.
 \label{eq:LB}
 \end{equation}
This Lagrangian leads to the Lorentz force dynamical equation
\[
\frac{d{\bi p}}{dt}=e\,{\bi u}\times{\bi B}.
\]

According to (\ref{eq:metricaEM}) and calling $K=eBmc$, the Finsler metric of spacetime is, 
\[
g_{00}=m^2c^2+\frac{Kxu^2u_y}{(c^2-u^2)^{3/2}},\quad
g_{11}=-m^2c^2+\frac{Kxu_y}{(c^2-u^2)^{3/2}}\left(c^2-u_y^2-u_z^2\right),
\]
\[
g_{22}=-m^2c^2+e^2B^2x^2+\frac{Kxu_y}{(c^2-u^2)^{3/2}}\left(3c^2-3u_x^2-2u_y^2-3u_z^2\right),
\]
\[
g_{33}=-m^2c^2+\frac{Kxu_y}{(c^2-u^2)^{3/2}}\left(c^2-u_x^2-u_y^2\right),
\]
\[
g_{01}=-\frac{Kxcu_x u_y}{(c^2-u^2)^{3/2}},\quad g_{02}=-\frac{Kxc}{(c^2-u^2)^{3/2}}(c^2-u_x^2-u_z^2),\]
\[
g_{03}=-\frac{Kxcu_y u_z}{(c^2-u^2)^{3/2}},\quad
g_{12}=\frac{Kxu_x}{(c^2-u^2)^{3/2}}\left(c^2-u_x^2-u_z^2\right),\]
\[
g_{13}=\frac{Kx}{(c^2-u^2)^{3/2}}u_x u_y u_z,\quad
g_{23}=\frac{Kxu_z}{(c^2-u^2)^{3/2}}\left(c^2-u_x^2-u_z^2\right),
\]
As we see, the metric coefficients are functions of the point, i.e., of the variable $x$, 
but they are functions of the three components of the velocity of the particle
$u_x,u_y,u_z$, and thus $g_{\mu\nu}(x,\dot{x})$. If the velocity is negligible with respect to $c$, the 
metric coefficients become
\[
g_{00}=m^2c^2,\quad
g_{11}=-m^2c^2,\quad
g_{22}=-m^2c^2+e^2B^2x^2,\quad
g_{33}=-m^2c^2,
\]
vanishing the remaining ones, and since the dependence on the velocity
has dissapeared the metric has been transformed into a Riemannian metric. 
Spacetime metric is Riemannian in the low velocity limit. 

These metric coefficients give rise to a restricted Lagrangian $\widetilde{L}_R$, 
 \begin{equation}
\widetilde{L}_R^2=m^2c^2(c^2\dot{t}^2-\dot{\bi r}^2)+e^2B^2x^2\dot{y}^2,
 \label{eq:LR}
 \end{equation}
such that when compared with (\ref{eq:LB}) we have an additional term
\[
\widetilde{L}_B^2=\widetilde{L}_R^2-2emcBx\dot{y}\sqrt{c^2\dot{t}^2-\dot{\bi r}^2}.
\]
From the restricted Lagrangian (\ref{eq:LR}) the force acting on the point particle is 
not longer the Lorentz force.

In the case (c) in a uniform gravitational field ${\bi g}$, the dynamical equations 
$d{\bi p}/dt=m{\bi g}$, independent of the mass of the particle, come from the Lagrangian
 \begin{equation}
\widetilde{L}_g=\widetilde{L}_0+m{\bi g}\cdot{\bi r}\dot{t}.
 \label{eq:Lg}
 \end{equation}
From the geodesic point of view it corresponds to an evolution in a spacetime with the Finslerian metric given by:
\[
g_{00}=m^2c^2+m^2({\bi g}\cdot{\bi r})^2/c^2-\frac{m^2c({\bi g}\cdot{\bi r})}{(c^2-u^2)^{3/2}}(2c^2-3u^2),\quad 
\]
\[
g_{11}=-m^2c^2+\frac{m^2c({\bi g}\cdot{\bi r})}{(c^2-u^2)^{3/2}}(c^2-u_y^2-u_z^2),
\]
\[
g_{22}=-m^2c^2+\frac{m^2c({\bi g}\cdot{\bi r})}{(c^2-u^2)^{3/2}}(c^2-u_x^2-u_z^2),
\]
\[
g_{33}=-m^2c^2+\frac{m^2c({\bi g}\cdot{\bi r})}{(c^2-u^2)^{3/2}}(c^2-u_x^2-u_y^2),
\]
\[
g_{01}=-\frac{m^2u^2({\bi g}\cdot{\bi r})}{(c^2-u^2)^{3/2}}u_x,\quad
g_{02}=-\frac{m^2u^2({\bi g}\cdot{\bi r})}{(c^2-u^2)^{3/2}}u_y,\quad
g_{03}=-\frac{m^2u^2({\bi g}\cdot{\bi r})}{(c^2-u^2)^{3/2}}u_z,
\]
\[
g_{12}=\frac{m^2c({\bi g}\cdot{\bi r})}{(c^2-u^2)^{3/2}}u_xu_y,\quad
g_{23}=\frac{m^2c({\bi g}\cdot{\bi r})}{(c^2-u^2)^{3/2}}u_yu_z,\quad
g_{13}=\frac{m^2c({\bi g}\cdot{\bi r})}{(c^2-u^2)^{3/2}}u_xu_z.
\]
If again, the velocity is negligible with respect to $c$, the nonvanishing coefficients are
\[
g_{00}=m^2c^2+m^2({\bi g}\cdot{\bi r})^2/c^2-2{m^2({\bi g}\cdot{\bi r})},\quad 
g_{11}=-m^2c^2+{m^2({\bi g}\cdot{\bi r})},\]
\[
g_{22}=-m^2c^2+{m^2({\bi g}\cdot{\bi r})},\quad
g_{33}=-m^2c^2+{m^2({\bi g}\cdot{\bi r})}.
\]
i.e.,
\[
g_{00}=m^2c^2\left(1-\frac{{\bi g}\cdot{\bi r}}{c^2}\right)^2, \quad g_{ii}=-m^2c^2\left(1-\frac{{\bi g}\cdot{\bi r}}{c^2}\right),\quad i=1,2,3,
\]
where the $g_{00}$ component is the same as the corresponding component of the Rindler metric 
corresponding to a noninertial accelerated observer or to the presence of a global uniform
gravitational field, in General Relativity.

A final example is the relativistic point particle in the Newtonian potential of a point mass $M$. 
The dynamical equations
\[
\frac{d{\bi p}}{dt}=-\frac{GmM}{r^3}{\bi r},
\]
independent of the mass of the particle, come from the
Lagrangian
 \begin{equation}
\widetilde{L}_N=\widetilde{L}_0+\frac{GmM}{cr}c\dot{t}.
 \label{eq:LN}
 \end{equation}
If we take into account (\ref{eq:metrica}) the metric is
\[
g_{00}=m^2c^2+\frac{G^2m^2M^2}{c^2r^2}-\frac{Gm^2Mc}{r(c^2-u^2)^{3/2}}(2c^2-3u^2),
\]
\[
g_{11}=-m^2c^2+\frac{Gm^2Mc^3}{r(c^2-u^2)^{3/2}}-\frac{Gm^2Mc(u_y^2+u_z^2)}{r(c^2-u^2)^{3/2}},
\]
\[
g_{22}=-m^2c^2+\frac{Gm^2Mc^3}{r(c^2-u^2)^{3/2}}-\frac{Gm^2Mc(u_x^2+u_z^2)}{r(c^2-u^2)^{3/2}},
\]
\[
g_{33}=-m^2c^2+\frac{Gm^2Mc^3}{r(c^2-u^2)^{3/2}}-\frac{Gm^2Mc(u_x^2+u_y^2)}{r(c^2-u^2)^{3/2}},
\]
\[
g_{01}=-\frac{Gm^2Mu^2u_x}{r(c^2-u^2)^{3/2}},\quad g_{02}=-\frac{Gm^2Mu^2u_y}{r(c^2-u^2)^{3/2}},\quad g_{03}=-\frac{Gm^2Mu^2u_z}{r(c^2-u^2)^{3/2}},
\]
\[
g_{12}=\frac{Gm^2M c u_xu_y}{r(c^2-u^2)^{3/2}},\quad
g_{23}=\frac{Gm^2M c u_yu_z}{r(c^2-u^2)^{3/2}},\quad
g_{31}=\frac{Gm^2M c u_zu_x}{r(c^2-u^2)^{3/2}},
\]
It is a Finsler metric, which in the case of low velocities only the diagonal components survive
\[
g_{00}=m^2c^2\left(1-\frac{2GM}{c^2r}+\frac{G^2M^2}{c^4r^2}\right)=m^2c^2\left(1-\frac{GM}{c^2r}\right)^2.
\]
If the last term  $G^2M^2/r^2c^4$ is considered negligible, it is the $g_{00}$ coefficient of
Schwarzschild's metric. 
The 
\[
g_{ii}=-m^2c^2\left(1-\frac{GM}{c^2r}\right),\quad i=1,2,3,
\] 
are different than in the Schwarzschild case. In any case we see that the
coefficients, in the low velocity limit, they reduce to the gravitational potential of the central mass $M$,
divided by $c^2$.

In all the examples, the free Lagrangian $\widetilde{L}_0$ of the spinless particle, has been transformed
by the interactions in the way
 \begin{equation}
\widetilde{L}_0^2=m^2c^2\eta_{\mu\nu}\dot{x}^\mu\dot{x}^\nu\quad\Rightarrow\quad \widetilde{L}^2=g_{\mu\nu}(x,\dot{x})\dot{x}^\mu\dot{x}^\nu,
 \label{eq:EMl}
 \end{equation}
where the new metric $g_{\mu\nu}(x,\dot{x})$ is a Finslerian metric. The low velocity limit
of the above metrics produce a Riemannian approximation which does not give rise to the
usual dynamical equations.

However, General Relativity states that gravity modifies the metric of spacetime producing
a new (pseudo-)Riemannian metric $g_{\mu\nu}(x)$, which is related through Einstein's equations
to the energy momentum distribution $T^{\mu\nu}$ of all forms of matter and energy.
The motion of a point particle in this gravitational background is a geodesic on spacetime, 
and therefore can be treated
as a Lagrangian dynamical problem with a Lagrangian
 \begin{equation}
\widetilde{L}_g^2=g_{\mu\nu}(x)\dot{x}^\mu\dot{x}^\nu.
 \label{eq:Grav}
 \end{equation}
In the spirit of unification of all interactions, in particular electromagnetism produces a Finsler metric of spacetime,
one is tempted to extend the formulation
of gravity (\ref{eq:Grav}) to (\ref{eq:EMl}) by allowing to the metric produced by gravity
to be also a function of the derivatives. Otherwise, to assume only a Riemannian metric
is to consider that gravity produces a different geometrization than any other interaction. 
In a region where the gravitational field can be considered uniform, the Lagrangian dynamics of the point
particle is equivalent to a geodesic problem in that region where the metric is 
necessarily a Finsler metric, as seen in the above examples. 
The elimination of the presence
of the velocities in the metric coefficients could be interpreted as a low velocity limit
of a more general gravitational theory.

\section{The spin structure of elementary particles}
\label{sec:spin}

In the mentioned reference \cite{Rivasbook} and previous works cited in, an elementary particle is defined
as a mechanical system whose kinematical space $X$ is necessarily a homogeneous space of the Poincar\'e group ${\cal P}$.
The idea is that an elementary particle cannot be divided and, if not annihilated with its antiparticle, 
cannot be deformed so that any state is 
just a kinematical modification of any one of them \cite{atomic}. 
When the initial state $x_1$ is modified by the dynamics, the subsequent states $x(\tau)$
can always be obtained from it by some change $g\in{\cal P}$ of inertial observer $x=gx_1$ and also $x_2=gx_1$,
so that given any two points $x_1,x_2\in X$ we can always find some $g\in{\cal P}$ which links them.
It is clear that the point particle manifold is a homogeneous space of ${\cal P}$ and thus 
fulfills this requirement, 
but it describes a spinless object and 
there are no spinless elementary particles in nature. To describe spin we have to enlarge 
this kinematical space
with the above constraint to obtain the largest homogeneous space of ${\cal P}$ 
to describe the elementary particle with the more complex structure.
The classical system that when quantized satisfies Dirac's equation corresponds
to a kinematical space spanned by the following variables $x\equiv(t,{\bi r},{\bi u},\balpha)$, 
which are interpreted as the time $t$, position of the center of charge ${\bi r}$, 
velocity of the center of charge ${\bi u}=d{\bi r}/dt$ at the speed 
of light $u=c$, and the orientation $\balpha$ of a cartesian system located at point ${\bi r}$  \cite{quantum}. 
It is a nine-dimensional
kinematical space described by four noncompact variables $(t,{\bi r})$ and five compact ones $(\theta,\phi,\balpha)$,
being $\theta,\phi$ the orientation of the velocity vector ${\bi u}$ and the orientation of the particle local frame $\balpha$.
The particle has a center of mass ${\bi q}$ which is expressed in terms of ${\bi r}$ and its time derivatives. Elementary
spinning particles have two distinguished points the center of mass and the center of charge 
which are different points.
The cartesian frame can be taken as the Frenet-Serret triad, so that the angular velocity of the particle can also be expressed
in terms of the derivatives of the position of the point ${\bi r}$.

The free motion of the center of charge ${\bi r}$ corresponds to a helix of constant curvature 
and torsion when expressed in terms of the Frenet-Serret triad, and at a velocity of constant absolute value $c$.

This classical model of elementary particle can be applied for leptons and quarks if, as assumed, they satisfy
in the quantum formalism Dirac's equation.

If we want to include gravitation we have to admit arbitrary changes of spacetime coordinates, not only
those given by the Poincar\'e group. This will produce a modification of the metric of the spacetime
submanifold, but also the modification of the remaining components of the metric on the whole kinematical $X$-space.
Because all known baryonic and leptonic matter and the gauge bosons are spinning objects we cannot start the geometrization
of matter by assuming that it is only the metric of the kinematical space of the point particle which is modified, because
there are no spinless objects in nature. We have to geometrize the kinematical space of the spinning particle
accordingly. We cannot make things simpler.

\section{Conclusions}

We consider that General Relativity is a constrained, and therefore a 
simpler formalism for describing gravity for two reasons: One is that the geometrization
of spacetime has to be enlarged to consider Finsler metrics instead of pseudo-Riemannian metrics, and another that
the manifold which describes the boundary states of spinning matter is larger than spacetime.

The manifold $X$ of the boundary variables of any Lagrangian dynamical system is always a Finsler metric space,
so that any variational approach is equivalent to a geodesic statement on this metric manifold.
This metric, which is in general a function of the variables $x\in X$ and their
derivatives $\dot{x}$, depends on the interactions, and to assume that gravitation only produces a modification
of the metric which is only a function of the point $x$, is a restriction of a more general formalism which allows
for this modification, in the spirit of unification of all interactions. 

The second constraint of a gravitational theory is that it has to be applied to a manifold larger than
spacetime, because spacetime is the boundary manifold of the spinless point particle and spinless elementary particles
seem not to exist in nature.

Without any assumption about general covariance, or any assumption about the dynamical behaviour
of spacetime, but in a Lagrangian framework, 
we have analysed several examples of 
the Finsler space structure of spacetime under different interactions. 
In all of them the new metrics
are true Finsler metrics which in the case of low velocity limit, and therefore a metric 
independent of the velocities, resemble the metrics obtained in a general relativity formalism.

It is possible that the spin structure of matter plays no role in the gravitational analysis of the solar system
and the usual treatment in terms of only spacetime variables is sufficient to describe planetary motions. 
But in cosmological models, when the velocity of particles is not negligible, redshits of order 6 and higher
have been quoted for several galaxies which correspond to velocities of $0.9c$, a metric dependent of the velocities would
produce a different analysis than a Riemannian one.
When a quantum analysis is required, may be in a neutron star with the magnetic moments of the particles
aligned, in a gravitational colapse with a huge density matter where gravitational effects associated to 
the spin structure are expected, is unavoidable to enlarge the formalism
to more general aspects of the form considered here in which the space $X$ is larger than spacetime and 
the metric should depend also on the velocities. The physical restrictions have to applied in the analysis
of the particular cases, not at the very begining of the formalism.

\ack{This work has been partially supported by 
Universidad del Pa\'{\i}s Vasco/Euskal Herriko Unibertsitatea grant  9/UPV00172.310-14456/2002.}


\end{document}